# Annealing, acid, and alcoholic beverage effects on $Fe_{1+y}Te_{0.6}Se_{0.4}$

Y. Sun[1,2], T. Taen[1], Y. Tsuchiya[1], Z. X. Shi[2], T. Tamegai[1,2]

[1]*Department of Applied Physics, The University of Tokyo, 7-3-1 Hongo, Bunkyo-ku, Tokyo 113-8656, Japan*

[2]*Department of Physics, Southeast University, Nanjing 211189, People's Republic of China*

## *Abstract*

We have systematically investigated and compared different methods to induce superconductivity in iron chalcogenide $Fe_{1+y}Te_{0.6}Se_{0.4}$ including annealing in vacuum, $N_2$, $O_2$, $I_2$ atmosphere, and immersing samples into acid and alcoholic beverages. Vacuum and $N_2$ annealing are proved to be ineffective to induce superconductivity in $Fe_{1+y}Te_{0.6}Se_{0.4}$ single crystal. $O_2$ and $I_2$ annealing, acid and alcoholic beverages can induce superconductivity by oxidizing the excess Fe in the sample. Superconductivity in $O_2$ annealed sample is in bulk nature, while $I_2$, acid and alcoholic beverages can only induce superconductivity near the surface. By comparing different effects of $O_2$, $I_2$, acid and alcoholic beverages, we propose a scenario to explain how the superconductivity is induced in the non-superconducting as-grown $Fe_{1+y}Te_{0.6}Se_{0.4}$.

# 1. INTRODUCTION

The recent discovery of superconductivity at 26 K in an iron oxypnictide LaFeAs(O, F) [1] has stimulated great interests among the condensed-matter physics community. Tremendous amount of works have been carried out, leading to the emergence of novel iron-based superconductor families with different crystal structures: 1111 (LaFeAs(O,F)) [1], 122 ((Ba,K)Fe$_2$As$_2$) [2], 111 (LiFeAs) [3] and 11 (Fe(Te,Se)) [4]. Among these, FeTe$_{1-x}$Se$_x$ has received special attention due to its simple crystal structure, which are composed of only Fe(Te,Se) layers. Superconductivity and magnetism of this system are not only dependent on the doping level, but also sensitive to Fe non-stoichiometry, which originates from the partial occupation of excess Fe at the interstitial site in the Te/Se layer [5, 6]. For the undoped parent compound Fe$_{1+y}$Te, the commensurate antiferromagnetic order can be tuned by the excess Fe to an incommensurate magnetic structure [5]. In Se-doped FeTe samples, excess Fe was found to suppress superconductivity and cause the magnetic correlations [6]. In the other end member Fe$_{1+y}$Se, superconductivity is reported to reside just in a very narrow concentration region of excess Fe [7]. To probe the intrinsic properties of FeTe$_{1-x}$Se$_x$ without the influence of excess Fe, some previous works have been performed to remove the effect of excess Fe by annealing it in different conditions. Superconductivity was found to be induced by annealing in vacuum [8, 9], air [10], oxygen [11, 12], nitrogen [13], I$_2$ vapor [14], and even immersing into nitric acid HNO$_3$ [13]. Recently superconductivity was also reported to be induced in FeTe$_{0.8}$S$_{0.2}$ by immersing it into alcoholic beverages [15, 16]. The interpretations for these treatments to induce superconductivity are still controversial, including the improvement of homogeneity [8], oxygen intercalation [11], and deintercalation of excess Fe [10, 13, 14]. Although superconductivity can be successfully induced, some annealed samples show a very broad transition width [11, 13, 14, 16], which implies an inhomogeneous superconductivity. In order to address these issues, and try to understand how the superconductivity is induced in Fe$_{1+y}$Te$_{0.6}$Se$_{0.4}$, we have systematically investigated different methods to induce superconductivity by annealing in vacuum, N$_2$, O$_2$, I$_2$ atmosphere, and immersing the samples into acid and alcoholic beverages. By comparing the results, we give a possible scenario to explain how the superconductivity is induced in the non-superconducting as-grown $Fe_{1+y}Te_{0.6}Se_{0.4}$.

## 2. EXPERIMENTAL METHODS

Single crystals with nominal compositions $FeTe_{0.6}Se_{0.4}$ and $FeTe_{0.8}S_{0.2}$ were prepared from high purity Fe (99.99%), Te (99.999%), Se (99.999%) and S (99.9999%) grains [8]. More than 10 g of stoichiometric quantities were loaded into a small quartz tube with $d_1 \sim 10$ mm$\phi$, evacuated, and sealed. Then we sealed this tube into a second evacuated quartz tube with $d_2 \sim 20$ mm$\phi$. The whole assembly was heated up to 1070 ℃ and kept for 36 h, followed by slow cooling down to 710 ℃ at a rate of 6 ℃/h. After that, the temperature was cooled down to room temperature by shutting down the furnace. The as-grown single crystal with typical dimensions of $5.0 \times 5.0 \times 1.0$ mm$^3$ for $FeTe_{0.6}Se_{0.4}$, and $2.0 \times 2.0 \times 0.1$ mm$^3$ for $FeTe_{0.8}S_{0.2}$ can be easily cleaved perpendicular to the $c$ axis. For vacuum annealing, sample was loaded into a quartz tube, which was carefully baked and examined that no appreciable amount of gas was emitted under the same condition as the sample annealing. The quartz tube was carefully evacuated by a diffusion pump before sealing the sample. For annealing with $I_2$, the sample was loaded into a quartz tube with $I_2$ chips (99.9%). For annealing in $O_2$ and $N_2$ atmosphere, after evacuating, we filled controlled amounts of $O_2$ and $N_2$ gas before sealing the quartz tube. During these processes, a diaphragm-type manometer with an accuracy of $10^{-3}$ torr was used for real-time monitoring the pressure in the system to prevent gas leakage and to check the amount of gas in the tube. Then the samples were annealed at 400℃ for more than one day, followed by water quenching. For $O_2$ annealing, we have tried annealing the sample under 0.1%, 1% and 1 atmosphere (atm) of $O_2$ gas. $Fe_{1+y}Te_{0.6}Se_{0.4}$ samples annealed both under 0.1 % and 1 % atm of $O_2$ show superconducting transition temperature, $T_c$, higher than 14 K, and the former exhibits larger $J_c$. By contrast, the sample annealed in 1 atm $O_2$ was totally damaged, which indicates that the sample itself was oxidized during annealing in too much $O_2$. Other pieces of as-grown samples were put into glass bottles (10 ml) filled with 20% hydrochloric acid HCl, beer (Asahi breweries Ltd.), red wine (Asahi breweries Ltd.), Japanese sake (Hakutsure Sake Brewing, Co. Ltd.), shochu (Iwagawa Jozo Co. Ltd.) or whisky (Suntory Holding Ltd.). The samples immersed into alcoholic beverages (beer, red wine, Japanese sake, shochu and whisky) were kept at 70 ℃ for 40 h. The sample immersed into 20% HCl was kept at room temperature for 100 h, because we found heating the sample up to 70 ℃ with acid damaged the sample quickly. Magnetization measurements on the as-grown and post-treated samples were performed using a commercial SQUID magnetometer (MPMS-XL5, Quantum Design). Magneto-optical (MO) images were obtained by using the local field-dependent Faraday effect in the in-plane magnetized garnet indicator film employing a differential method [17, 18].

## 3. RESULTS AND DISCUSSION

Figure 1 (a) shows temperature dependence of zero-field-cooled (ZFC) and field-cooled (FC) magnetization at 5 Oe for the as-grown, vacuum, $N_2$, $O_2$, and $I_2$ atmosphere annealed $Fe_{1+y}Te_{0.6}Se_{0.4}$ single crystals. The as-grown crystals usually show no superconductivity or very weak diamagnetic signal below 3 K. These results show that superconductivity cannot be induced by vacuum or $N_2$ atmosphere annealing, which is quite different from the previous reports [8, 9, 13]. Annealing in $I_2$ atmosphere can successfully induce superconductivity, while the $T_c$ is lower and the transition width is much broader than that in the crystal annealed in 0.1% atm of $O_2$, which shows a $T_c$ higher than 14 K with the transition width less than 1 K (obtained from the criteria of 10 % and 90 % of the magnetization result at 2 K). We also tested annealing crystals in poor vacuum (~ 1 torr), which also enhanced the superconductivity. Thus, the previously reported vacuum or $N_2$ atmosphere annealing induced superconductivity is probably due to the poor vacuum or air leakage during the annealing process [8, 9, 13]. Figure 1 (b) shows the temperature dependent magnetization of $Fe_{1+y}Te_{0.6}Se_{0.4}$ crystals immersed into alcoholic beverages (beer, red wine, Japanese sake, shochu, whisky) and 20% HCl, together with the result of $O_2$ annealing for comparison. All these crystals exhibit superconductivity, among which the crystal immersed into beer has the largest diamagnetic signal. Although superconductivity can be induced by alcoholic beverages and acid, similar to the crystal annealed with $I_2$, $T_c$ is lower and the transition width is much broader than that annealed in $O_2$ atmosphere. The alcoholic beverage induced superconductivity is also reported in S-doped FeTe sample [16]. For comparison, temperature dependent magnetization of $Fe_{1+y}Te_{0.8}S_{0.2}$ single crystal immersed into beer at 70 ℃ for 1 week is also plotted in Figure 1 (b), which shows a $T_c$ about 7 K close to the previous report [16]. In this case, the transition width is also very broad, similar to the alcoholic beverage effect on $Fe_{1+y}Te_{0.6}Se_{0.4}$ crystals, and the diamagnetic signal is even smaller than that in $Fe_{1+y}Te_{0.6}Se_{0.4}$ treated by all of the methods.

It is well known that even if the sample is mostly non-superconducting, the diamagnetic signal becomes significant when the non-superconducting region is covered by superconducting region. Thus, to further confirm the nature of superconductivity, magnetic hysteresis loops (MHLs) of samples immersed into alcoholic beverages, acid as well as annealed in $I_2$ atmosphere were measured at 2 K and shown in Figure 2 (a). Although obvious superconducting hysteresis loop can be witnessed, it is too weak to persist up to applied fields larger than 20 kOe. The absolute value of magnetization is also very small compared with that of $O_2$ annealed $Fe_{1+y}Te_{0.6}Se_{0.4}$ crystals, as shown in Figure 2 (b).

From the MHLs, we can obtain the critical current density, $J_c$, by using the Bean model [19] :

$$J_c = 20\frac{\Delta M}{a(1-a/3b)}, \qquad (1)$$

where $\Delta M$ is $M_{down} - M_{up}$, $M_{up}$ and $M_{down}$ are the magnetization when sweeping fields up and down, respectively, $a$ and $b$ are sample widths ($a < b$). Magnetic field dependence of $J_c$ at 2 K is summarized in Figure 3. $J_c$ of the $O_2$ annealed sample reaches the value larger than $5\times 10^5$ A/cm$^2$ at zero field. It is robust under applied field, keeping a value larger than $1\times 10^5$ A/cm$^2$ even at 50 kOe. Although the value of $J_c$ is still lower than that of $Ba(Fe_{1-x}Co_x)_2As_2$ single crystal [20-22], it is the largest among those reported in Fe(Te,Se) [8, 23-28]. This fact demonstrates the high quality of our $O_2$ annealed sample. To further confirm the homogeneous current flow within the sample, we took MO images of 0.1% atm $O_2$ annealed crystal in the remanent state. In the inset of Figure 3, a typical MO image taken at 6 K after cycling the field up to 800 Oe along the $c$ axis is shown. The MO image manifests a typical roof-top pattern, similar to that observed in high quality $Ba(Fe_{1-x}Co_x)_2As_2$ single crystals [20, 29, 30], indicating a nearly uniform current flow in the crystal. The uniformity of $J_c$ is much better than that reported in ref. [8]. Profile of the magnetic induction along the dashed line in the MO image is also shown in the inset of Figure 3. $J_c$ can be roughly estimated by $J_c \sim \Delta B / t$, where $\Delta B$ is the trapped field and $t$ is the thickness of the sample. With $\Delta B \sim 1076$ G and $t = 50$ μm, $J_c$ is estimated as $2.2 \times 10^5$ A/cm$^2$ at 6 K, consistent with estimation from the Bean model. The large $J_c$ estimated from MHLs and MO image cannot be sustained only by the superconductivity near the surface. Thus, the superconductivity induced by $O_2$ annealing must be a bulk property of the sample. In $Fe_{1+y}Te_{0.6}Se_{0.4}$ single crystal immersed into alcoholic beverages, acid as well as annealing with $I_2$, $J_c$ at 2 K under zero field is smaller than $1 \times 10^4$ A/cm$^2$. In addition, $J_c$ decreases very quickly with increasing field, and becomes smaller than $1 \times 10^2$ A/cm$^2$ at fields above 20 kOe. For $FeTe_{0.8}S_{0.2}$, the value of the zero-field $J_c$ is smaller than 500 A/cm$^2$, and easily suppressed by the modest applied field, which decreases to lower than 100 A/cm$^2$ at fields about 10 kOe. The much smaller value of $J_c$ compared with that in the $O_2$ annealed sample manifests that the superconductivity comes from the surface.

To directly prove that the superconductivity induced by $I_2$, acid and alcoholic beverages is only near the surface, we cut the four edges of the sample, polished the double surfaces (more than half of the sample was removed after polishing), and then compared the superconducting property of the whole sample and the inner part. In Figures 4 (a) and (b), the results on samples annealed in 1% atm $O_2$ and immersed into beer are compared. In the case of $O_2$ annealed sample, $T_c$ of the inner part is almost the same as that in the whole sample, and the diamagnetic signal as well as the $J_c$ do not decrease but a little larger than the whole sample. It should be noted that the 1% atm $O_2$ annealed sample shows smaller $J_c$ than that annealed under 0.1% atm $O_2$, although $T_c$ is similar. The decrease in $J_c$ may be caused by the fact that the sample itself was partially oxidized by the excess $O_2$, especially on the surface, which changes color after annealing as reported before [10]. After removing the over-oxidized surface, the diamagnetic signal and $J_c$ are increased. By contrast, in the sample immersed into beer, both the diamagnetic signal and the $J_c$ of the inner part are decreased to about 10 % of the whole sample. The significant degradation in superconducting properties after removing the surface directly proves that the observed superconductivity mainly comes from the surface.

Compiling obtained results, we propose a scenario to explain how the superconductivity is induced by different methods. Vacuum and $N_2$ annealing cannot induce superconductivity, which indicates that homogeneity is not the key factor for inducing superconductivity. The previously reported vacuum and $N_2$ annealing induced superconductivity may be caused by the leakage of $O_2$, since we found only trace amount of $O_2$ can successfully enhance superconductivity. Furthermore all oxidizing agents, $O_2$, $I_2$, acid, and alcoholic beverages can induce superconductivity possibly by removing the excess Fe in the sample rather than the $O_2$ itself doped into the crystal. Among these, only $O_2$ can easily intercalate between the layers of the single crystal, and deintercalates the excess Fe from the inner part of the sample because of its small size, inducing bulk superconductivity in $Fe_{1+y}Te_{1-x}Se_x$. On the other hand, $I_2$, acid and the alcoholic beverages can just induce superconductivity near the surface because of their relatively larger size preventing them from intercalating between the layers of the crystal. Thus, the rate of excess Fe deintercalation is much slower compared with the $O_2$ annealing, and can induce superconductivity only near the surface of the crystal. Very recently, the alcoholic beverage effect was also reported in $Fe_{1+y}Te_{0.9}Se_{0.1}$ polycrystal [31].

The $O_2$ annealing and alcoholic beverages have been reported to be effective in $Fe_{1+y}Te_{1-x}S_x$ single crystals [15, 32]. While for $O_2$ annealed $Fe_{1+y}Te_{1-x}S_x$ single crystals, superconductivity can be suppressed by the following vacuum annealing, which is quite different from the case of $O_2$ annealed $Fe_{1+y}Te_{1-x}Se_x$, in which the superconductivity is stable under vacuum annealing (data not shown). It suggests that the oxygen is doped into $Fe_{1+y}Te_{1-x}S_x$ rather than deintercalating the excess Fe. However, although superconductivity is successfully induced in $Fe_{1+y}Te_{1-x}S_x$, there are still no proofs for bulk superconductivity like large $J_c$ or obvious jump of specific heat at $T_c$. More efforts along this line are necessary to clarify the origin of different effects between $Fe_{1+y}Te_{1-x}Se_x$ and $Fe_{1+y}Te_{1-x}S_x$ single crystals.

## 4. CONCLUSION

In conclusion, we have found that vacuum and $N_2$ annealing cannot enhance superconductivity in $Fe_{1+y}Te_{0.6}Se_{0.4}$ single crystal. $O_2$ and $I_2$ annealing, acid and alcoholic beverages can induce superconductivity by oxidizing the excess Fe in the sample. Large value of $J_c \sim 5 \times 10^5$ A/cm$^2$ obtained from MHLs and MO image shows that bulk superconductivity can be induced by $O_2$ annealing. The self-field $J_c$ at 2 K of the samples annealed in $I_2$ atmosphere, immersed into acid and alcoholic beverages are smaller than $1 \times 10^4$ A/cm$^2$, and suppressed by the modest magnetic field. Furthermore, the diamagnetic signal and $J_c$ of the inner part of these samples are much smaller than those obtained from the whole sample. These results indicate that the $I_2$ atmosphere, acid and alcoholic beverages can only induce superconductivity near the surface.

## 5. Acknowledgments

This work is partly supported by the Natural Science Foundation of China, the Ministry of Science and Technology of China (973 project: No. 2011CBA00105).

# Figure caption

Figure 1: Temperature dependence of zero-field-cooled (ZFC) and field-cooled (FC) magnetization at 5 Oe for (a) the as-grown, vacuum, $N_2$, 0.1% atm $O_2$, and $I_2$ atmosphere annealed $Fe_{1+y}Te_{0.6}Se_{0.4}$. Because of the low $T_c$ and weak diamagnetic signal, ZFC and FC curves of the as-grown, vacuum and $N_2$ annealed samples are almost merged together as pointed out by the arrow. (b) $Fe_{1+y}Te_{0.6}Se_{0.4}$ immersed into alcoholic beverages (beer, red wine, Japanese sake, shochu, whisky), 20% HCl, and $FeTe_{0.8}S_{0.2}$ immersed into beer, together with the result of $O_2$ annealing for comparison.

Figure 2: (a) Magnetic hysteresis loops (MHLs) of $Fe_{1+y}Te_{0.6}Se_{0.4}$ immersed into alcoholic beverages, acid, annealed in $I_2$ atmosphere and $FeTe_{0.8}S_{0.2}$ immersed into beer measured at 2 K. (b) MHLs of samples immersed into alcoholic beverages, acid, annealing in $I_2$ atmosphere plotted together with the 0.1% atm $O_2$ annealed sample for comparison.

Figure 3: Field dependence of $J_c$ at 2 K for $Fe_{1+y}Te_{0.6}Se_{0.4}$ annealed in $I_2$, 0.1% atm $O_2$, immersed into alcoholic beverages, acid, and $FeTe_{0.8}S_{0.2}$ immersed into beer. The inset shows an MO image in the remanent state taken at 6 K in 0.1 % atm $O_2$ annealed $Fe_{1+y}Te_{0.6}Se_{0.4}$. The lower inset shows the magnetic induction profile along the dashed line in the MO image.

Figure 4: Temperature dependence of magnetization of the whole and inner part of (a) 1% atm $O_2$ annealed and (b) beer treated $Fe_{1+y}Te_{0.6}Se_{0.4}$. The inset shows the magnetic field dependent critical current density, $J_c$, of the whole and inner part of each crystal.

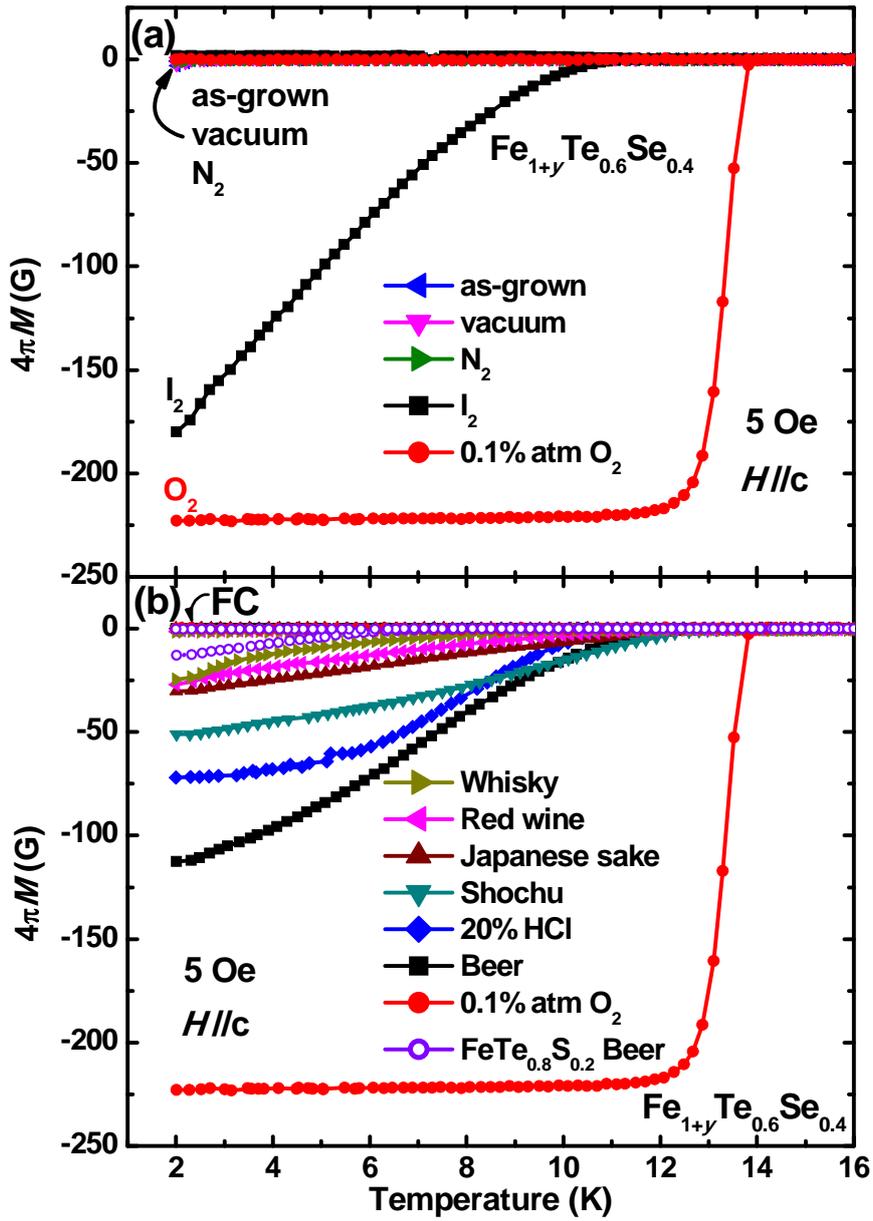

**Figure. 1** Y. Sun *et al.*

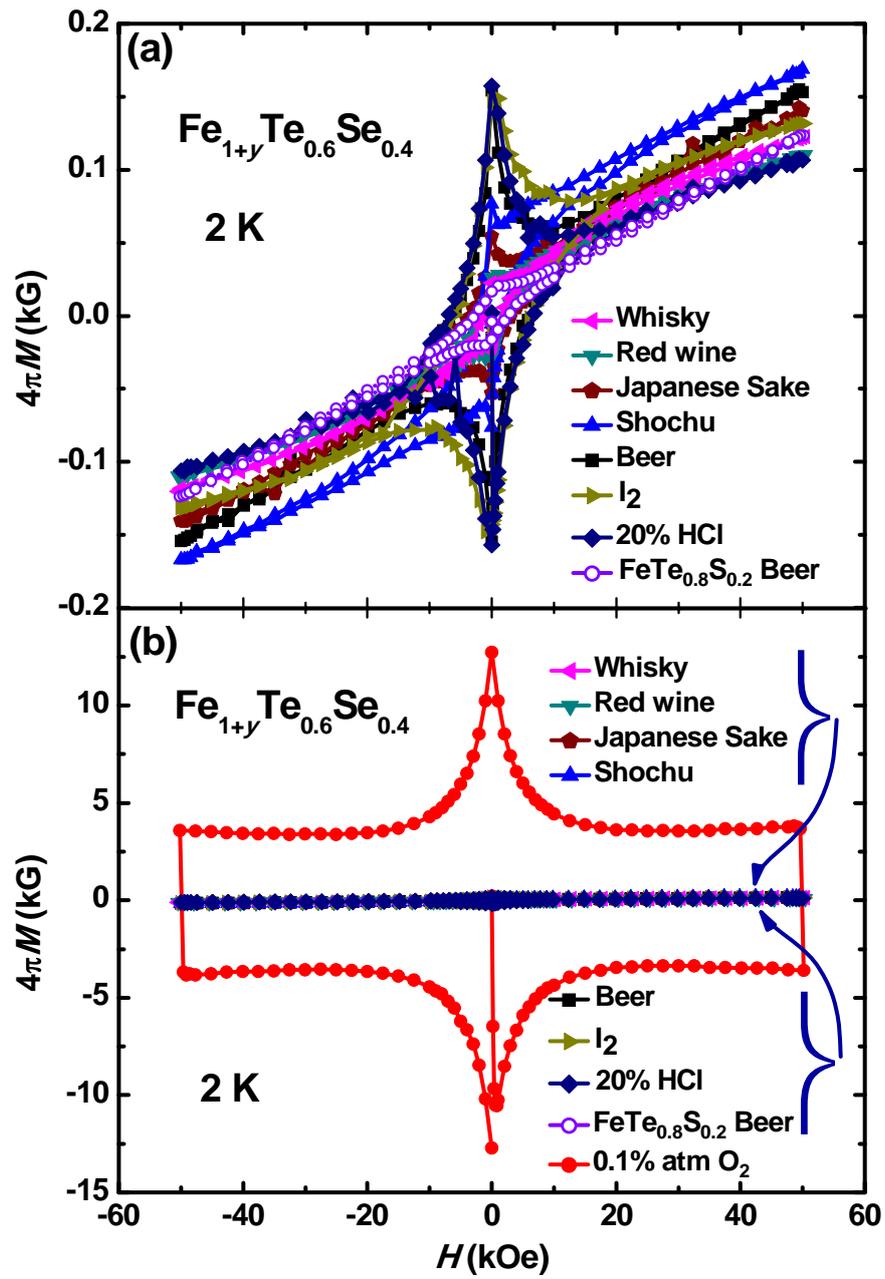

Figure. 2 Y. Sun *et al.*

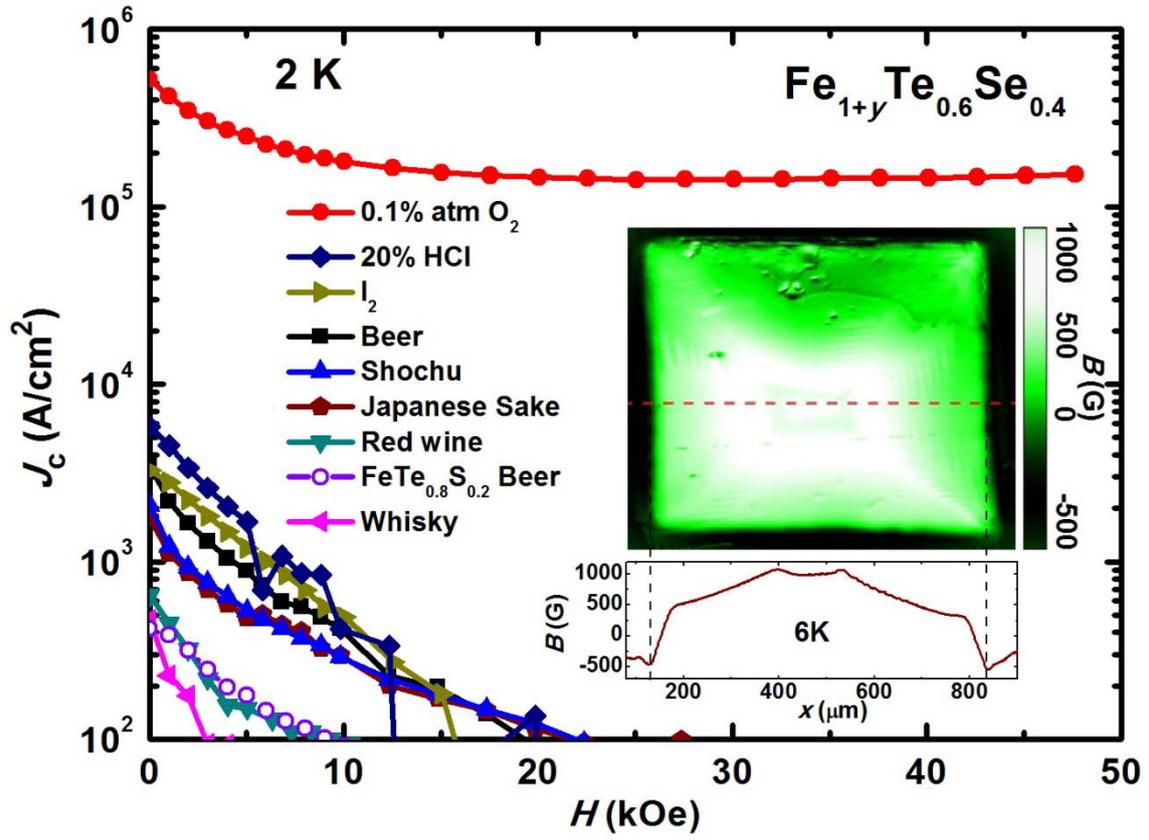

**Figure. 3 Y. Sun** *et al.*

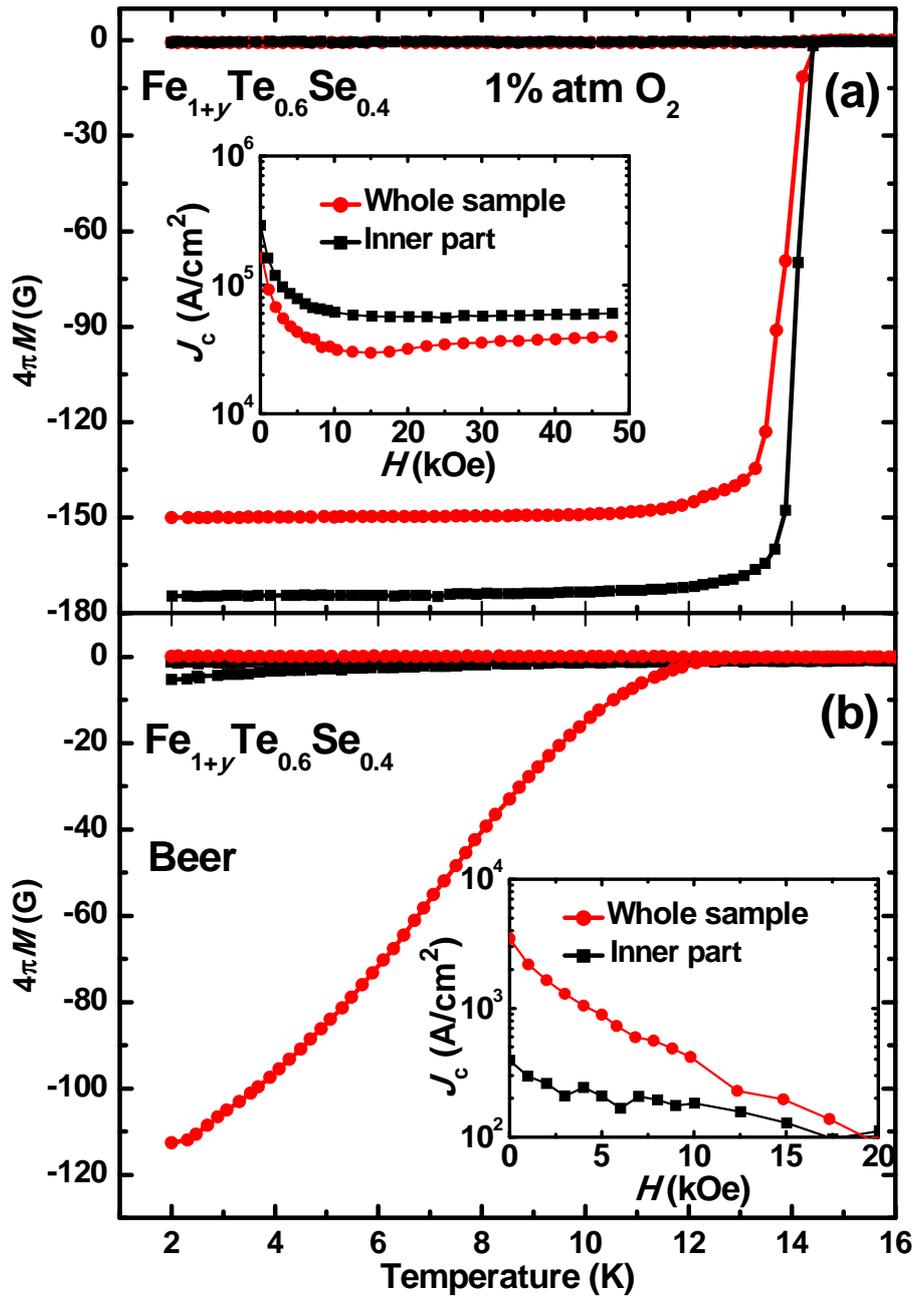

Figure. 4 Y. Sun *et al.*